# Enhancement of perpendicular magnetic anisotropy and its electric field-induced change through interface engineering in Cr/Fe/MgO

A. Kozioł-Rachwał*[1,2], T. Nozaki[1], K. Freindl[3], J. Korecki[2,3], S. Yuasa[1] and Y. Suzuki[1,4]

[1] National Institute of Advanced Industrial Science and Technology, Spintronics Research Center, Tsukuba, Ibaraki 305-8568, Japan

[2] Faculty of Physics and Applied Computer Science, AGH University of Science and Technology, al. Mickiewicza 30, 30-059 Kraków, Poland

[3] Jerzy Haber Institute of Catalysis and Surface Chemistry, Polish Academy of Sciences, ul. Niezapominajek 8, 30-239 Kraków, Poland

[4] Graduate School of Engineering Science, Osaka University, 1-3 Machikaneyama, Toyonaka, Osaka 560-8531, Japan

Recently, perpendicular magnetic anisotropy (PMA) and its voltage control (VC) was demonstrated for Cr/Fe/MgO (Physical Review Applied 5, 044006 (2016)). In this study, we shed a light on the origin of large voltage-induced anisotropy change in Cr/Fe/MgO. Analysis of the chemical structure of Cr/Fe/MgO revealed the existence of Cr atoms in the proximity of the Fe/MgO interface, which can affect both magnetic anisotropy (MA) and its VC. We showed that PMA and its VC can be enhanced by controlled Cr doping at the Fe/MgO interface. For Cr/Fe (5.9 Å)/Cr (0.7 Å)/MgO with an effective PMA of 0.8 MJ/m$^3$, a maximum value of the voltage-controlled magnetic anisotropy (VCMA) effect of 370 fJ/Vm was demonstrated.

**Introduction**

Electric field-controlled magnetic effects have recently attracted attention due to their potential application in high density, low power-consumption memory. Various studies on voltage-controlled magnetic effects have been reported for different materials, including ferromagnetic (FM) semiconductors, FM metals, or multiferroics (for review see Ref.[1] and Ref. within). For FM metals the VC of magnetism via electrical modulation of magnetic anisotropy (MA),[2,3] the Curie temperature,[4] domain wall motion,[5,6,7] and the interfacial asymmetric exchange interaction[8] have been demonstrated so far. Moreover, recent reports on bistable magnetization switching[9,10,11] and excitation of high frequency magnetization dynamics in an ultrathin FeCo layer by using the VC of MA[12,13] have contributed to the realization of a new class of voltage-controlled magnetoresistive random access memory (MRAM).[14]

For FM metal films, in contrast to semiconductors,[15] the voltage-induced changes in magnetism are restricted to the atoms closest to the surface due to the short screening length. Several mechanisms have been proposed to be responsible for the voltage-controlled magnetic anisotropy (VCMA) in FM systems. Theoretical studies attributed the origin of the VCMA effect to the spin-dependent screening of the electric field in FM layers[16] or the change in the relative occupancy of the orbitals of atoms at the interface related to the accumulation/depletion of electrons.[17,18,19] Experimentally, changes in perpendicular magnetic anisotropy (PMA) under an electric field ranges between tenths to thousands of fJ/Vm. The biggest variations of the areal density of the anisotropy energy under an electric field (the so-called VCMA coefficient) were demonstrated for systems in which voltage-induced redox reactions,[20] charge trapping effects,[21] electromigration,[22] or magnetostriction[23] were involved. Although the above-mentioned effects can contribute to the VCMA coefficients being as high as a few thousand fJ/Vm, a practical application of the systems for which VCMA is related to chemical reactions at the interface or magnetostriction is limited due to the low operating speed and poor write endurance. However, from a practical point of view, the systems for which the VCMA effect mainly originates from modifications of the electronic structure seems to be promising due to the high-speed operation. Aside from a large VCMA effect, a large PMA is desired to satisfy high thermal stability at reduced dimensions. Since both the VCMA effect and PMA usually have interfacial origins, the growth of high quality and controllable metal/dielectric interfaces is crucial for the understanding and further development of systems that exhibit voltage-tunable MA.

Recently, high interfacial MA[24,25,26] and a large VCMA effect[27] were found in an ultrathin Fe layer embedded between Cr and MgO. For an Fe layer with nominal thickness greater than 6 Å a high interfacial anisotropy of 2.1mJ/m$^2$ and a linear VCMA coefficient of about 100fJ/Vm were observed. For Fe layers thinner than 6 Å, a considerable reduction in PMA and the saturation magnetization ($M_s$) were noted. Several effects can be responsible for this behavior. One possible explanation is an intermixing between Fe and Cr,[28] which can occur especially when an annealing treatment is used during the deposition process.[29] Interestingly, for the thickness of the Fe layer for which a reduction in PMA and $M_s$ were noted, a large VCMA coefficient was observed. For a nominal thickness of the Fe layer of 5.3 Å, a VCMA coefficient of 290fJ/Vm was obtained under a negative voltage, while the PMA remained almost unchanged under a positive voltage. These results suggest that for a deeper understanding of the origin of PMA and its VC, the chemical structure of Cr/Fe/MgO should be analyzed in detail.

In the first part of this paper, we discuss an experimental verification of the local atomic structure of the Cr/$^{57}$Fe/MgO with the use of conversion electron Mössbauer spectroscopy (CEMS). We show the effects of annealing after the deposition of the MgO barrier on the magnetic properties and chemical structure in the system. We demonstrate that although no heavy intermixing exists between Fe and Cr in Cr/Fe/MgO, Cr impurities exist at the Fe/MgO interface after annealing. These Cr atoms in the proximity of the Fe/MgO interface can influence the PMA and VCMA.

Inspired by the CEMS results, in the second part of the paper we investigated the role of the insertion of an ultrathin Cr layer at the Fe/MgO interface on PMA and its voltage-induced change in Cr/Fe ($t_{Fe}$)/Cr ($d_{Cr}$)/MgO. We demonstrate that a light doping of Cr at the Fe/MgO interface enhances both the PMA and the VCMA coefficient. A maximum value of the VCMA of 370 fJ/Vm was obtained for an Fe layer thickness of 5.9 Å and a Cr layer thickness of 0.7Å inserted between the Fe and MgO layers.

**Experimental details**

Fully epitaxial MgO (10 Å)/Cr (300 Å)/Fe (*t*)/Cr (*d*)/MgO (25 Å)/Fe (100 Å)/Ta (50 Å)/Ru (70 Å) multilayers were grown on polished MgO(001) substrates under ultrahigh vacuum conditions. An MgO (30 Å)/Cr (300 Å)/$^{57}$Fe (6 Å)/MgO (25 Å) heterostructure was grown for the Mössbauer spectroscopy studies, and no Cr doping was used at the Fe/MgO interface. After the MgO substrate was annealed, a Cr buffer layer was grown at 200°C and annealed at 800°C. A wedge-shaped Fe layer with a thickness $t_{Fe}$ ranging from 3 Å to 7 Å was grown at 150°C and annealed at 250°C for 20 minutes. A sharpening of the reflection high-energy electron diffraction (RHEED) patterns was observed after annealing the thin Fe layer, which revealed an improvement of the crystalline quality as well as a smoothening of the annealed Fe surface. The Fe wedge was capped with a wedge-shaped Cr layer with a thickness $d_{Cr}$ ranging from 0 Å to 2 Å and with the Cr wedge gradient perpendicular to that of Fe (see Fig. 1). No change in the RHEED pattern was observed after deposition of the Cr-submonolayer on top of the Fe layer. Following Cr deposition, an MgO barrier was grown at room temperature and annealed at 350°C. Finally, the top Fe layer was deposited and capped with the Ta/Ru bilayer prepared by sputtering. Squared pillars were fabricated on this sample using optical lithography, ion beam etching, and the lift-off process. As a result, a matrix of elements with different Fe and Cr thicknesses was formed. The tunneling magnetoresistance (TMR) was measured in the current-perpendicular-to-plane geometry (CPP) using the standard two-probe method under an in-plane external magnetic field on junctions with different Fe and Cr thicknesses. During a measurement, an in-plane magnetic field was applied parallel to the Fe [001] direction. For a fixed Fe thickness, we noted a gradual increase in the resistance-area (RA) product and a decrease in TMR along with an increase in Cr thickness at the Fe/MgO interface; for example, RA = 250 k$\Omega\mu m^2$ and TMR = 26% were noted for Cr/Fe/MgO at $t_{Fe}$ = 5.9 Å, while RA = 312 k$\Omega\mu m^2$ and TMR = 13% were measured when 0.5 Å of Cr was inserted between Fe and MgO. Note that in the studied Fe thickness regime ($t_{Fe}$ = (3-7) Å), the magnetization of the bottom Fe layer was aligned along the surface normal, while a direction of magnetization of the top Fe layer was aligned in-plane along the easy-axis Fe [001] direction. Thus, the above-mentioned TMR values correspond to the relative orthogonal alignment between magnetizations.

Conversion electron Mössbauer spectroscopy (CEMS) measurements were performed for Cr/$^{57}$Fe(6Å)/MgO using a standard Mössbauer spectrometer equipped with a He/CH$_4$-flow proportional detector and a $^{57}$Co (Rh) source. The CEMS spectra were collected in the normal incidence geometry and fitted using commercial software. A Voight-line based method was applied to fit the spectra, in which the distribution of the hyperfine parameters is represented by the sum of the Gaussian components.[30]

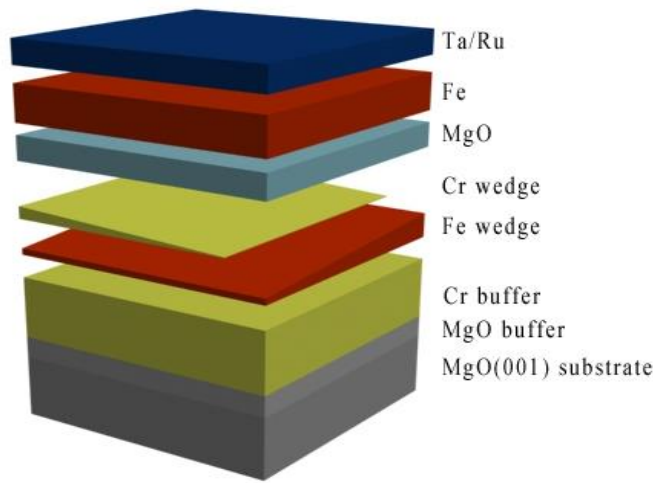

*Figure 1 Schematic drawing of the sample.*

**Results**

**I.   Cr/Fe/MgO structure probed by Mössbauer spectroscopy**

As mentioned in the introduction, a post-deposition annealing of the Fe layer improves its surface quality. The role of the second annealing (after deposition of MgO) is to improve the crystallinity of MgO; however, it can affect the chemical and magnetic structure of Cr/Fe/MgO, especially at the interfaces. To verify the influence of the second annealing on the magnetic properties and chemical composition of Cr/Fe/MgO, a sample before annealing and a sample after annealing were compared. Figure 2 presents the CEMS spectra of Cr/$^{57}$Fe (6 Å)/MgO collected for the sample before annealing (Fig. 2(a)) and the annealed sample after deposition of MgO (Fig. 2(b)). The nominal thickness of Fe corresponds to four atomic layers and, for an ideal film, one should expect a four-component spectrum reflecting Fe sites in different coordinations. Instead, both spectra revealed a rather complex hyperfine pattern, which indicates a static long-range magnetic order and deviation from an ideal layered structure. The spectra were deconvoluted into two groups of magnetic sub-spectra that are distinct by isomer shifts (IS) and their correlations with the hyperfine magnetic field ($B_{hf}$) distributions. The fit parameters are summarized in the Supplementary Information (Table 1). Sub-spectrum A (blue line) represents all of the Fe atoms except those at the Fe/MgO interface (nominally 3.4 atomic monolayers (ML)) and sub-spectra B and C (green and brown line respectively for the non-annealed and annealed samples) describe Fe atoms situated at the Fe/MgO interface, as judged based on the isomer shift value discussed below. Additionally, for the sample before annealing, a weak sub-spectrum P with small $B_{hf}$ (site P) was identified (pink line). The key fit parameter for the magnetic sub-spectra was the intensity ratio (R) of the second (or fifth) to the third (or fourth) line of the sextet components: $R = I_{2(5)}/I_{3(4)}$. The R value is determined by the angle θ between the hyperfine magnetic field (local magnetization) and the γ-ray direction:

$$R = 4sin^2\theta/(1 + cos^2\theta).$$

For the given CEMS geometry (in which the γ-rays are parallel to the surface normal), the perpendicular (θ = 0°) and in-plane (θ = 90°) magnetization results in R = 0 and R = 4, respectively.

Sub-spectrum A was decomposed into the smallest necessary discrete number of magnetic six-line Voigt components with the IS, $B_{hf}$, quadrupole interaction parameter (ε), and relative intensity as the fit parameters. The IS values of the components were linearly correlated with the hyperfine field, which is well justified for the Fe-Cr systems[31]. For both samples, the fits gave small ε values, which is typical for cubic symmetry in metals. Moreover, the ε is similar for all components in the non-annealed sample (ε = -0.026 ± 0.004) and is linearly correlated with $B_{hf}$, for the annealed sample, which signifies a higher degree of structural order. The resulting $B_{hf}$ distribution is well reproduced by a sum of Gaussian profiles whose width was fixed to 1.0 T for all components. The components identified for sub-spectrum A are attributed to Fe atoms with a different number of Cr neighbors. It was shown that for the Fe/Cr interface the $B_{hf}$ of Fe atoms could be expressed by the following formula:[32, 33]

$$B_{hf}(n_1, n_2) = B_{hf}(0,0) + n_1 B_1 + n_2 B_2 \quad (1),$$

where $B_{hf}(0,0)$ denotes the hyperfine field of bulk Fe, $n_1(n_2)$ is the number of the nearest (next nearest) Cr neighbors, and $B_1 = 3.19$ T ($B_2 = 2.15$ T) is the contribution to the hyperfine magnetic field from one Cr atom in the first (second) shell around the Fe atom. Note that, because $B_{hf}$ for Fe is negative, the Cr neighbors lower the absolute value of $B_{hf}$ experienced by Fe atom, which means that higher Cr coordinations contribute to components with a smaller magnetic hyperfine splitting. Klinkhammer *et al.*[28] made a refinement of the formula (1) to account for an enhancement of the hyperfine field at the Fe/Cr interface. For ultrathin films, $B_1$ and $B_2$ may differ from the above values,[34] but the general trend that relates local coordination of Fe and $B_{hf}$ is preserved.

To summarize, for an ideally flat and sharp Cr/Fe interface, only the hyperfine fields of atoms situated within the first and second layer from the Cr/Fe interface should be affected by the Cr proximity. Thus, only three components should be identified in the sub-spectrum A which contributes to about 85% of the total spectral intensity (nominally representing 3.5 ML): $B_{hf}(4,1)$ for the interface layer, $B_{hf}(0,1)$ for the layer next to the interface, and $B_{hf}(0,0)$ for deeper Fe sites. Instead, sub-spectrum A shows a multimodal $B_{hf}$ distribution; 9 and 7 components were identified for the sample before and after annealing, respectively. The $B_{hf}$ distributions $P(B_{hf})$ are shown in Figure 3, together with the IS of the components. In agreement with previous reports,[28] we observed a positive linear $B_{hf}$ vs. IS correlation. This can be explained by an increase of s-electron density at the Fe nuclei with the increasing number of Cr neighbors.

Two groups of $B_{hf}$ are distinctly resolved for the non-annealed sample (Fig. 3a). The low $B_{hf}$ group is represented by components with $|B_{hf}| < 25$ T, and the high-$B_{hf}$ group, which peaked at $|B_{hf}| = 30.4$ T, contain components with $|B_{hf}| > 25$ T. The relative intensity of the components with $|B_{hf}| < 25$ T is 31%, which is equivalent to 1.3 ML of Fe. This group can be ascribed to the Fe atoms located in the vicinity of a diffused Fe/Cr interface. According to the Mössbauer studies of the Cr/Fe (40 Å)/Cr system, Fe atoms with the (4,1) configuration should be described by $B_{hf} = -23$ T for the sharp Cr/Fe interface.[28] Any deviation from the ideal, sharp interface, either configurational (steps or kinks) or chemical (interfacial Cr and Fe mixing), should lead to additional spectral components with $B_{hf}$ slightly lower or higher than the ideal interfacial value. For example, an Fe atom diffused in the Cr-interfacial layer changes the configuration of four Fe atoms in the interfacial Fe layer to (3,1) and one Cr atom in the interfacial Fe layer changes the configuration of four interfacial Fe atoms to (4,2). On the other hand, since the Fe layer in our experiment is much thinner than the 40 Å Fe layer referenced above, a finite size reduction of magnetization and $|B_{hf}|$ is obvious, we assign the component with a $|B_{hf}| = 20.7$ T to the atoms at the sharp interface with the (4,1) configuration (Fig. 4, grey), in agreement with previous observations for ultrathin Fe (001) sandwiched between Cr.[30] Other components with $|B_{hf}| < 25$ T are assigned to Fe atoms in the mixed interfacial Fe-Cr layers: $B_{hf} = 18.2$ T and 23.5 T could correspond to the (4,2) and (3,1) configurations, respectively, which are characteristic not only for the mixed Fe-Cr layer but also for Fe atoms at steps or kinks (Fig. 4, green). Additionally, Fe atoms within a Cr-rich atomic interfacial layer (Fig. 4, blue) are responsible for the components with $B_{hf} < 15$ T; however, due to low intensity, their interpretation is less reliable. Fe atoms diluted in Cr contribute to the weak sub-spectrum P. The linear IS vs. $B_{hf}$ correlation may not be valid for these Cr rich configurations, and therefore this contribution could not be included in the subspectrum A.

The second group of components with $|B_{hf}| > 25$ T constitutes 54% of the total spectral intensity, which is equivalent to 2.2 ML of Fe. The components with $|B_{hf}| = 26.4$ T, 28.2 T, 30.4 T, and 32.9 T are assigned to Fe configurations with $n_1 + n_2 = 3, 2, 1$, and 0, respectively, which can be found in the Fe layers next to the interface if some Cr atoms penetrate deeper into the Fe layer. (Fig. 4, orange). Finally, the component with the highest $|B_{hf}| = 35.1$ T and small positive IS relative to α-Fe is associated with Fe atoms in the sharp Fe/MgO interface (Fig. 4, black), where Fe retains its metallic character.[35] On the contrary, a small fraction of the Fe atoms, with strong bonds to the O atom in MgO (Fig. 4, violet), contribute to sub-spectrum B (Fig. 2(a), green), and are characterized by a distinctly positive IS = $(0.61 \pm 0.03)$ mm/s that identifies an oxidation state of Fe. Two similar forms of Fe were found in the previous CEMS studies of the Fe/MgO interfaces.[36]

The character of the CEMS spectrum changes after annealing (compare Fig. 3(a) and 3(b)). In sub-spectrum A, two groups of the magnetic components are not as pronounced as those of the non-annealed sample. The relative weight of components with $|B_{hf}| < 25$ T is almost the same as for the non-annealed sample (33%); however, the component with $|B_{hf}| < 15$ T) is no more present. The relative contribution of the component from Fe atoms in the (4,1) configuration (11%) is higher than that for the sample before annealing (8%). Furthermore, an average hyperfine magnetic field of the components with $|B_{hf}| < 25$ T was increased from 19.4 T before to 21.3 T after annealing, which suggests a sharpening of the Cr/Fe interface. Also, the components associated with Fe atoms located further from the Fe/Cr interface are modified after annealing. First, a bulk-like component that corresponds to the (0,0) configuration became more pronounced, and the metallic component with a high value of $|B_{hf}| = 35.1$ T (associated with a sharp Fe/MgO interface) as well as the sub-spectrum B of an oxidic character disappeared. Instead, sub-spectrum C with $B_{hf} = (-26 \pm 1.9)$ T and IS = (0.20 ± 0.03) mm/s and the relative contribution of 16.5%, was recognized in a spectrum measured for the annealed sample (Fig. 2, brown). We interpret this sub-spectrum as originating from an interface formed by a mixed Fe–Cr atomic layer and MgO. Formation of such a layer can be understood if Cr segregation occurs during annealing. According to a simple thermodynamic consideration, Cr should segregate to the surface due to the higher surface energy of Fe compared to that of Cr.[37] First principle calculations revealed the complexity of the segregation process and showed that the segregation energy depends on Cr concentration and the surface orientation.[38] In particular, it was shown that the segregation of Cr impurities is favorable for the Fe (001) surface for an optimal Cr concentration within the Fe layer, which well explains the present case.

Sub-spectrum C has replaced the oxide-like sub-spectrum B in the non-annealed sample. This means that the Cr atoms at the Fe/MgO interface prevent formation of the oxide-like Fe–MgO bonds observed for the non-annealed sample. Furthermore, the component with $|B_{hf}| = 35.1$ T, which is of a metallic character and is interpreted as originating from Fe atoms at the sharp Fe/MgO interface, disappeared from the CEMS spectrum after annealing of the sample. This result is understandable if the presence of Cr at the Fe/MgO interface is considered; i.e. the hyperfine parameters at the Fe–Cr/MgO interface combine the properties of the sharp Fe/MgO interface and the influence of neighboring Cr atoms that lead to a reduction of $|B_{hf}|$.

In short, the annealing has considerably changed the interfacial structure and the character of the Cr distribution in the Fe layer. Before annealing both interfaces were smeared and Cr tended to intermix and diffuse in the Fe layer. The annealing led to the configurational sharpening of the interfaces related to the Cr segregation and formation of a mixed Fe-Cr/MgO interface.

Together with the appearance of sub-spectrum C for the annealed sample, a striking change in the intensity ratio R was observed. While an in-plane magnetization alignment (R = 4) is recognized before annealing, the magnetization aligns with the perpendicular direction after annealing (R = 0). This suggests that a small Cr doping at the Fe/MgO interface is favorable for establishing the PMA in the Cr/Fe/MgO system.

This conclusion is significant in light of recent research on PMA and its VC in Cr/Fe/MgO.[26,27] As mentioned in the introduction, the maximum VCMA coefficient of about 290 fJ/Vm was observed for an Fe layer of nominal thickness of 5.3 Å, which is equivalent to 3.6 ML. This means that Cr impurities that segregate towards the Fe/MgO interface during annealing can be substantial for the MA and its VC in the system. To verify how a Cr insertion influences MA and its VC, systematic studies of the effective anisotropy and VCMA coefficient were performed for Cr/Fe (*t*)/Cr (*d*)/MgO as a function of different Fe and Cr thicknesses.

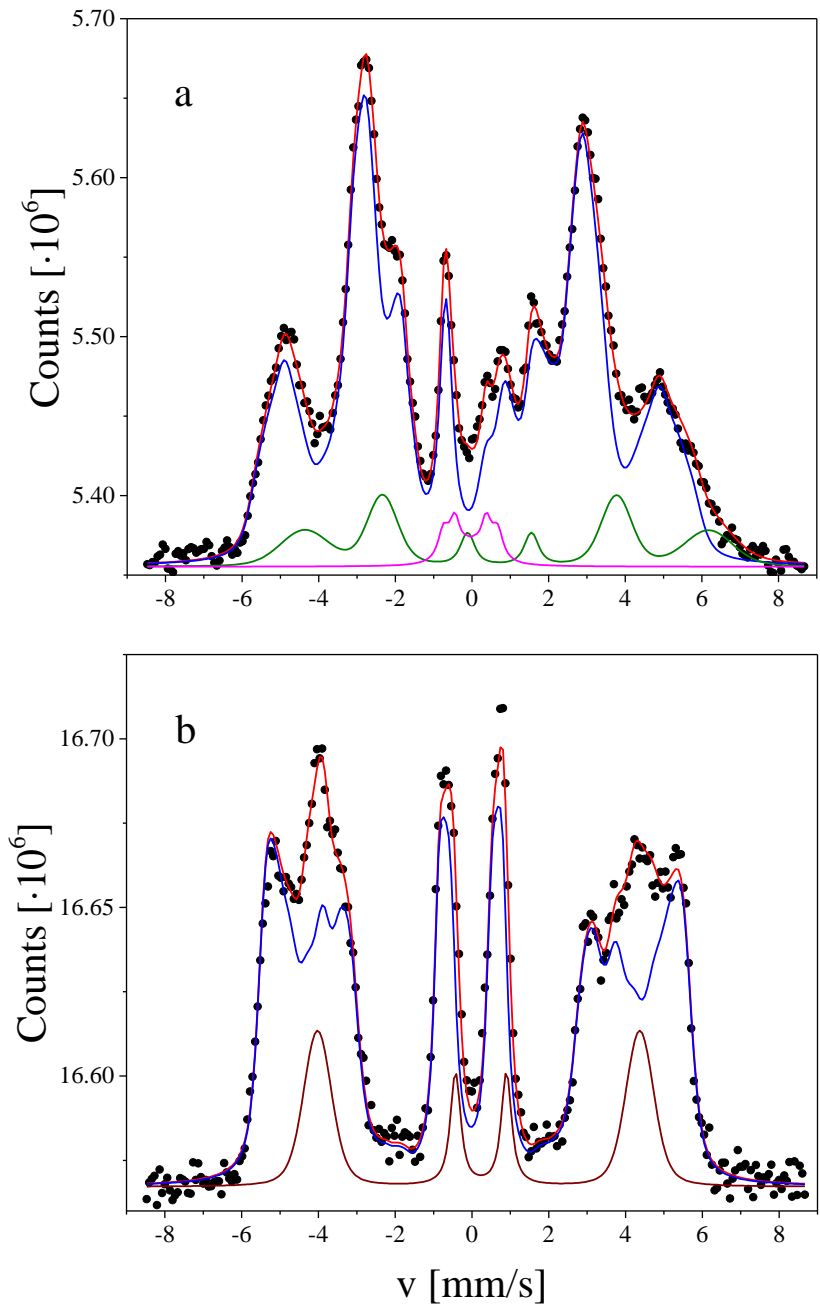

*Figure 2  CEMS spectrum of Cr/$^{57}$Fe/MgO collected after deposition of MgO (a) and after annealing at 300$^0$C (b).*

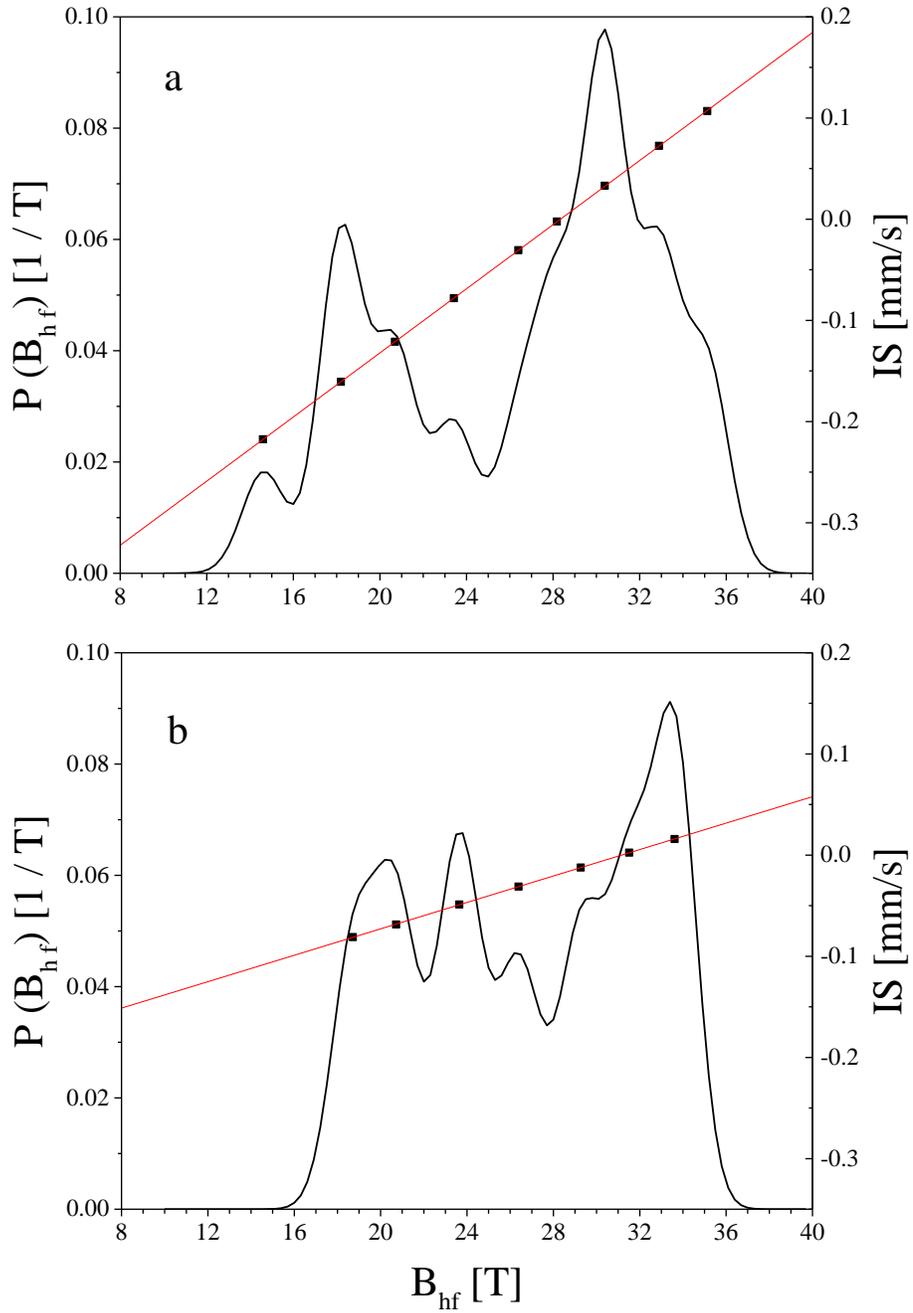

*Figure 3 Hyperfine field distribution (black line) and the isomer shift (black squares) for non-annealed (a) and annealed (b) sample, linearly correlated to the hyperfine filed distribution. The spectral intensity of the components and isomer shifts were shown on the left and right scale, respectively.*

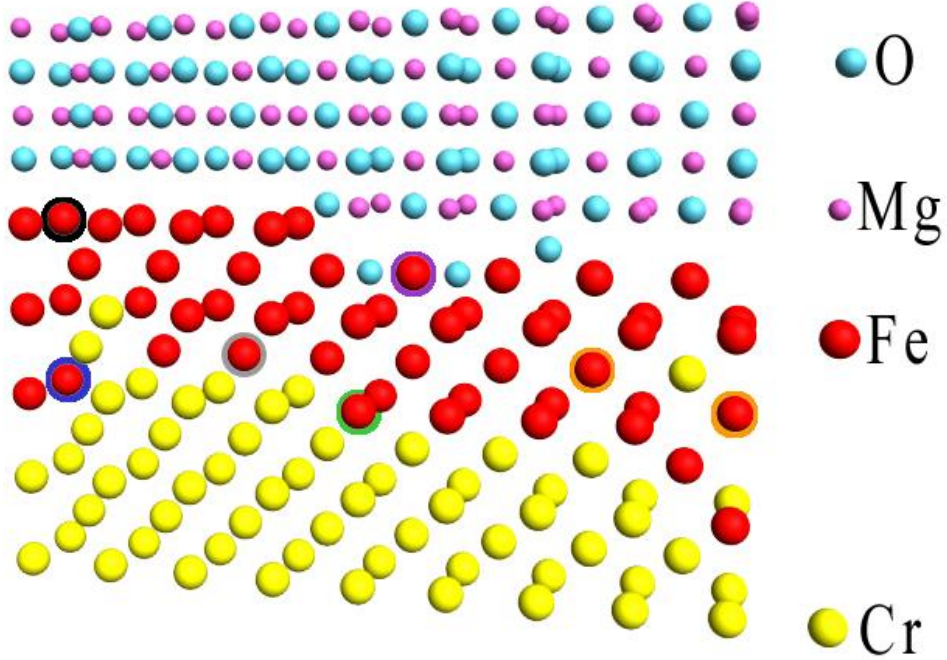

*Figure 4 Model of Cr/Fe/MgO structure of showing interface Fe atoms at the specific interfacial sites. Yellow, red, blue and pink balls represents Cr, Fe, O and Mg atoms respectively. Examples of interfacial sites determined from CEMS spectra were marked by circles with different colors.*

## II. Cr insertion dependence on PMA and its VC

The effective magnetic anisotropy constants for different Fe and Cr thicknesses were determined from magnetoresistance measurements by using the method described in Ref.[39]. Examples of normalized conductance curves, which represent the in-plane component of the Fe layer magnetization, are shown in Figs. 5(a) and (b) for an Fe thickness of 5.1 Å and 6.6 Å, respectively, for different Cr thicknesses. The shaded area in Fig. 5(b) (shown as an example for $d = 0$), when multiplied by $M_s$, is a measure of the PMA energy density. For an Fe thickness of 5.1 Å, we observed a gradual decrease of the saturation field ($H_s$) when an ultrathin Cr layer was inserted at the Fe/MgO interface (Fig. 5(a)). On the contrary, for thicker Fe layers initially, for 0.16 Å ≤ $d$ <

0.41 Å, we observed an increase in $H_s$ with increasing Cr thickness at the interface (Fig. 5(b)), whereas further increase of $d_{Cr}$ resulted in a decrease of $H_s$.

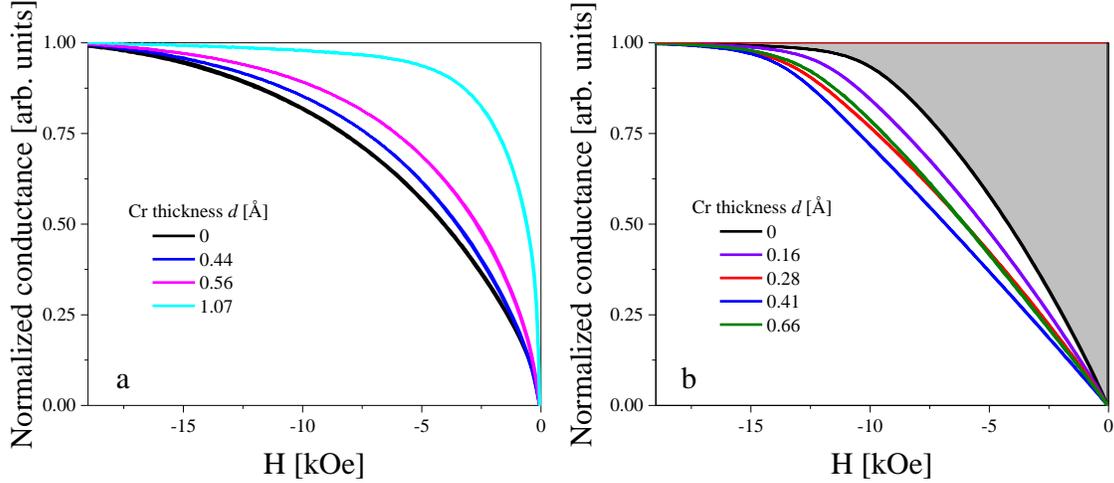

*Figure 5 Normalized tunneling conductance obtained for Fe thickness of 5.1Å (a) and 6.6 Å (b) for different Cr thicknesses.*

This suggests that while Cr doping at the Fe/MgO interface causes a reduction in PMA for very thin Fe layers ($t < 5.5$ Å), for thicker Fe films, it can contribute to an enhancement of PMA. Figure 6 summarizes a $K_{eff}$ dependence on Cr thickness obtained for different Fe thicknesses, where a positive $K_{eff}$ indicates an out-of-plane easy-axis of magnetization. The $M_s$ values, which are necessary for the estimation of $K_{eff}$ were obtained from SQUID measurements for selected Fe thicknesses. The $M_s$ values for intermediate Fe thicknesses were determined from areal magnetization (M/S) vs. $t_{Fe}$ dependence. For the samples with Cr doping at the Fe/MgO interface, we used (M/S)($t$) dependence obtained for different Fe thicknesses and a Cr layer inserted between Fe and MgO with a thickness of 0.5 Å (see: Supplementary Information).

Insertion of sub-monolayer Cr at the Fe/MgO interface induces substantial changes in the effective anisotropy in the Cr/Fe/MgO system. Whereas for $t < 5.5$ Å, even 0.5 Å of Cr at the interface causes a decrease of positive $K_{eff}$, sub-monolayer Cr doping for thicker Fe films leads to an enhancement of PMA. Furthermore, the Cr thickness for which a maximum value of $K_{eff}$ was observed is shifted towards higher $d_{Cr}$ values when the Fe thickness is increased. These results indicate that Cr doping at the Fe/MgO interface can be optimized for a maximum enhancement of PMA. For $t_{Fe}$ = 5.91 Å and $d_{Cr}$ = 0.28 Å or for $t_{Fe}$ = 6.61 Å and $d_{Cr}$ = 0.41 Å, we noted $K_{eff}$ of approximately 1.0 MJ/m$^3$.

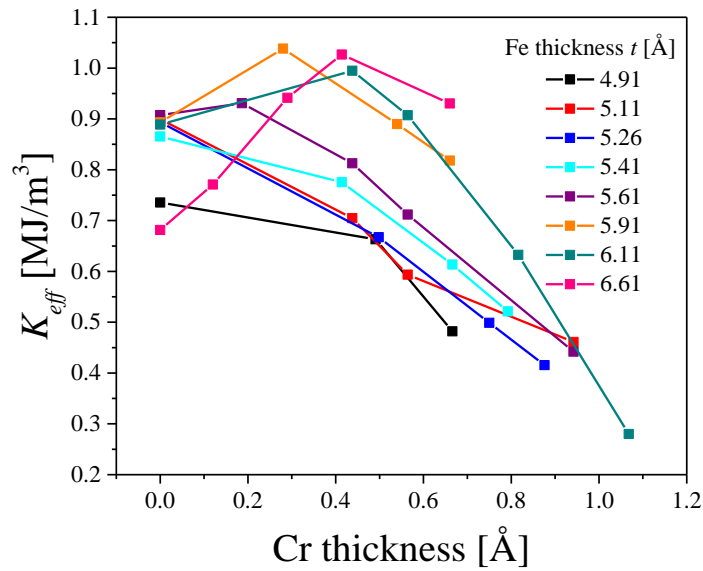

*Figure 6 $K_{eff}$ dependence on Cr thickness d for different Fe thicknesses.*

The VC of MA in the Cr/Fe/Cr/MgO system was studied by TMR measurements under different bias voltages, similar to our previous studies[27]. The effects of resistance and TMR ratio dependence on the bias voltage were eliminated through normalization of the TMR curves. The bias voltage was adjusted between the range -0.8 V < U < 0.8 V, which corresponds to the electric field range of -320 mV/nm < E < 320 mV/nm at the Fe/MgO interface. During measurements, we used a configuration in which a positive external voltage induces electron accumulation at the Fe/MgO interface. Figure 7 shows exemplary normalized conductance curves determined from the normalized TMR curves obtained for $t_{Fe}$ = 5.9Å and $d_{Cr}$ = 0.54Å under different voltages. For these Fe and Cr thicknesses, an electric field causes pronounced changes in the shape of the normalized conductance curve, which indicates a strong VCMA effect. Figure 8 presents a summary of the areal density of the anisotropy energy ($K_{eff}\,t$) dependence on the electric field ($E$) determined for Fe thicknesses of (a) 5.1 Å, (b) 5.4 Å, (c) 5.9 Å, (d) 6.1 Å for different Cr thicknesses inserted between Fe and MgO.

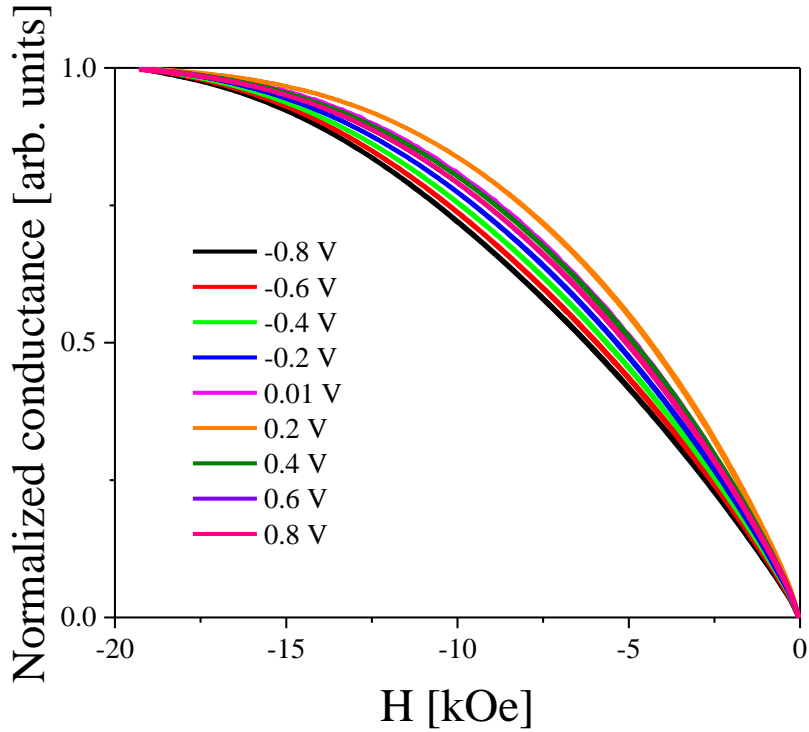

*Figure 7 Exemplary normalized conductance curves obtained for an Fe thickness of 5.9Å and Cr thickness of 0.54Å measured under different bias voltages.*

In agreement with our previous studies, for E < 100 mV/nm, we noted a linear increase of $K_{eff}\,t_{Fe}$ with an increase of negative electric field strenght. For the thinnest Fe layers ($t_{Fe}$ < 5.4 Å), a gradual reduction of the slope in the $K_{eff}\,t_{Fe}$ ($E$) dependence was observed together with increasing Cr thickness (compare black, blue and green data in Fig. 8(a)). In contrast, for thicker Fe layers, an insertion of Cr at the Fe/MgO interface results in an increase of the VCMA coefficient. An evident increase in the slope of the $K_{eff} t_{Fe} (E)$ dependence was noted for $t_{Fe} \geq 5.4$ Å for optimal Cr doping (Figs. 8(b)–(d)).

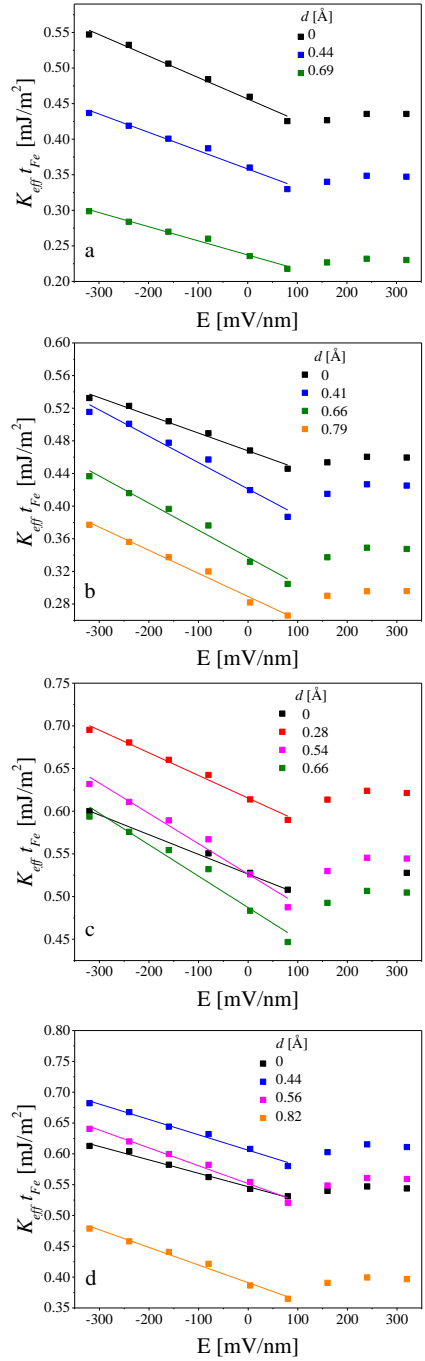

*Figure 8 $K_{eff}$ $t_{Fe}$ vs. electric field dependence obtained for the Fe thickness of 5.1Å(a), 5.4Å(b), 5.9Å(c) and 6.1Å(d) for different Cr thicknesses (d). The slopes of linear fits indicate VCMA coefficients. Summary of VCMA coefficients were shown on Fig. 9.*

A summary of the VCMA coefficients determined for different Fe and Cr thicknesses is shown in Fig. 9 (open squares) together with the $K_{eff}(d)$ dependence (filled circles). For the thinnest Fe layer of 5.1 Å, a gradual decrease of the VCMA coefficient with increasing Cr thickness was found, from 302 fJ/Vm for $d_{Cr} = 0$ Å to 257 fJ/Vm for $d_{Cr} = 0.44$ Å and finally to 199fJ/Vm for $d_{Cr} = 0.69$Å (Fig. 9, black squares). For thicker Fe layers, an enhancement of the VCMA coefficient was observed for sub-angstrom Cr doping. The strongest influence of Cr on the VCMA was obtained for $t_{Fe} = 5.61$ Å and $t_{Fe} = 5.91$ Å. For an Fe thickness of $t_{Fe} = 5.61$ Å, we noted a VCMA coefficient of 226 fJ/Vm without Cr insertion, which was enhanced to about 360 fJ/Vm with Cr doping of 0.56 Å (blue squares). For $t_{Fe} = 5.91$ Å, an enhancement of the VCMA coefficient from 230 fJ/Vm to 367 fJ/Vm was noted when a 0.66 Å-thick Cr layer was inserted at the Fe/MgO interface (Fig. 6, pink squares). For a given Fe thickness, the Cr thickness for which a maximum VCMA coefficient was noted is shifted towards thicker Cr in comparison with the thickness for which a maximum $K_{eff}$ was obtained. However, for some Fe thicknesses, a high $K_{eff}$ and enhanced VCMA coefficient could be obtained by using Cr insertion. The highest value of the VCMA coefficient of 367 fJ/Vm was noted for an Fe thickness of 5.9 Å and a Cr thickness of 0.7 Å, for which $K_{eff}$ was estimated to be 0.8 MJ/m$^3$.

As we shown in the experiment, an enhancement of $K_{eff}$ and the VCMA coefficient could be obtained with a small amount of Cr doping at the Fe/MgO interface. However, the origin of the effect is unclear. The most probable cause is the Cr impurities induce modification of the electronic structure of Fe. Modification of the band structure of Fe atoms at the Fe/MgO interface could affect both PMA and its VC. Another explanation assumes a change of a strain at the Fe/MgO interface due to the Cr doping. Although a big strain should not be expected in the system due to the small lattice mismatch between Fe and Cr (0.6%), recent ab-initio calculations showed that even small modification of the strain can dramatically affect on the VCMA.[40]

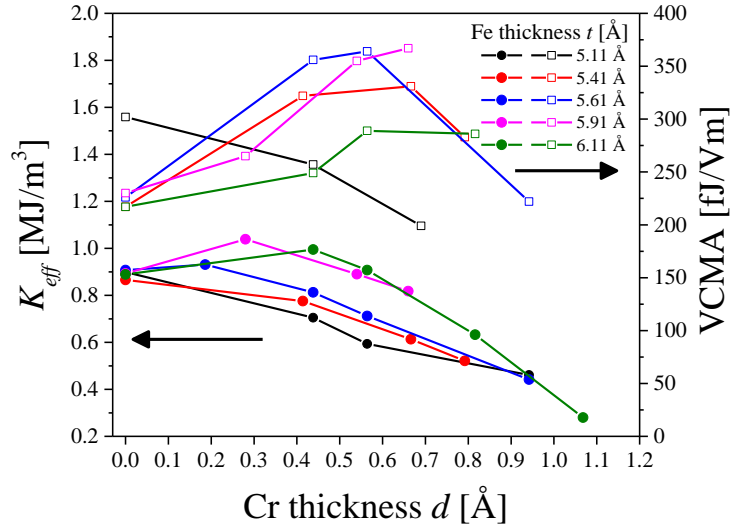

*Figure 9 Effective anisotropy (filled circles, left scale) and VCMA coefficient (open squares, right scale) dependence on Cr thickness for different Fe thicknesses. The lines are guides to the eye.*

**Conclusion**

In summary, we proved that sub-monolayer Cr insertion at the Fe/MgO interface could enhance both PMA and its electric field induced change. Similarly to our previous results, we noted an increase of PMA when a negative voltage was applied. Moreover, for $t_{Fe} \geq 5.4$ Å we showed that the change of $K_{eff}$ under an electric field was more pronounced when a Cr layer was inserted at the Fe/MgO interface. This result, together with a detailed analysis of the chemical structure of the Cr/Fe/MgO system with the use of Mössbauer spectroscopy, shows that the presence of Cr at the interface may induce a strong VCMA effect.

**Acknowledgements**

We thank E. Usuda for assistance with the experiments. This work was partly supported by ImPACT Program of Council for Science, Technology and Innovation, the Strategic AIST integrated R&D program, "IMPULSE", and a Grants-in-Aid for Scientific Research (26709046). K. F. and J. K. were supported in part by the National Science Center (NCN), Poland (Grant No. 2011/02/A/ST3/00150).

**Author Contributions**

A. K.-R. and T. N. designed the experiments and performed the sample fabrication, TMR measurements, and data analysis. S. Y. and Y. S. developed techniques for deposition, micro-fabrication, and measurements. K. F. and J. K. prepared samples for the Mössbauer spectroscopy (MS) studies and performed MS measurements and analysis. A. K.-R. wrote the manuscript with review and input from Y. S., J. K., and T. N. All authors contributed to the planning, discussion and analysis of this research.

# Supplementary information

*Table 1 Hyperfine parameters derived from the numerical fits of the CEMS spectra for Cr/$^{57}$Fe/MgO before and after annealing. IS denotes the average value of the isomer shift with respect to α-Fe, $B_{HF}$ is the average hyperfine magnetic field, $\Delta B_{HF}$ denotes the average Gaussian width of the $B_{HF}$ distribution for a given site, and QS is the average quadrupole splitting.*

| | | before annealing | | |
|---|---|---|---|---|
| **Component/ Subcomponent** | **IS[mm/s]** | **$B_{hf}$[T]/$\Delta B_{hf}$[T]** | **ε or QS [mm/s]** | **RI [%]** |
| A/1 | -0.21(3) | 14.6(3) | -0.026(4) | 4 |
| A/2 | -0.16(3) | 18.2(2) | -0.026(4) | 13 |
| A/3 | -0.12(3) | 20.7(5) | -0.026(4) | 8 |
| A/4 | -0.07(3) | 23.4(7) | -0.026(4) | 6 |
| A/5 | -0.03(3) | 26.3(9) | -0.026(4) | 5 |
| A/6 | 0.00(4) | 28(1) | -0.026(4) | 9 |
| A/7 | 0.04(4) | 30(2) | -0.026(4) | 20 |
| A/8 | 0.07(4) | 32.9(5) | -0.026(4) | 12 |
| A/9 | 0.11(4) | 35.1(3) | -0.026(4) | 8 |
| B | 0.61(4) | 32.7(2)/4.2(3) | 0.09(2) | 12 |
| P | 0.04(2) | 4.4(1) | 0 | 3 |
| | | after annealing | | |
| A/1 | -0.08(4) | 18.7(4) | -0.06(4) | 9 |
| A/2 | -0.07(4) | 20.7(6) | -0.05(4) | 11 |
| A/3 | -0.05(5) | 23.6(3) | -0.02(4) | 13 |
| A/4 | -0.03(5) | 26.4(6) | -0.01(4) | 9 |
| A/5 | -0.01(5) | 29.3(7) | 0.01(5) | 10 |
| A/6 | 0.00(5) | 31.5(8) | 0.03(5) | 12 |
| A/7 | 0.02(5) | 33.6(2) | 0.05(5) | 18 |
| B | 0.10(3) | 26.0(2)/1.8(2) | 0 | 17 |

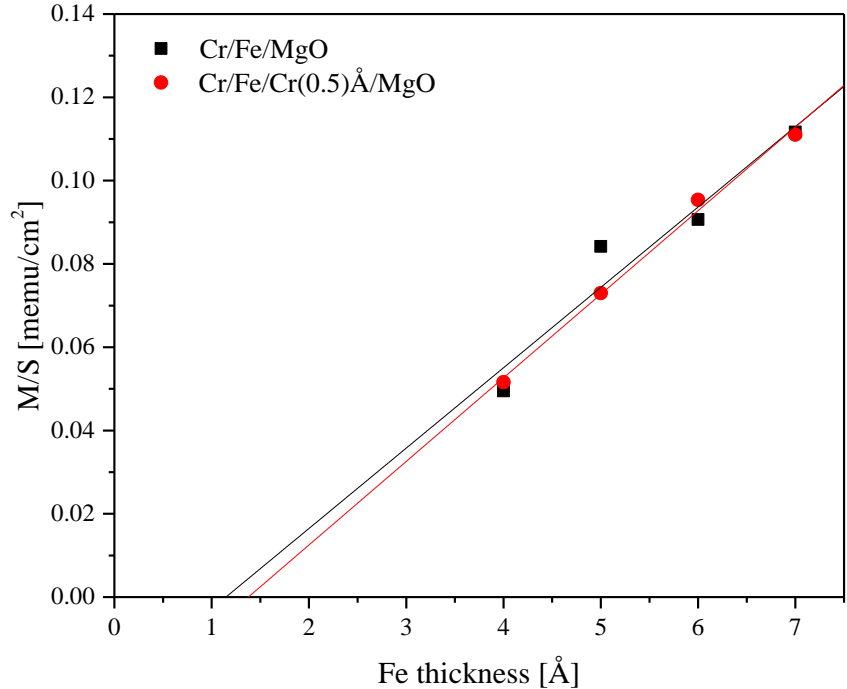

*Figure 1a Areal magnetization dependence on Fe thickness obtained from SQUID measurements for Cr/Fe/MgO (black squares), and Cr/Fe/Cr/MgO with a Cr thickness of 0.5Å (red squares).*